\documentclass[11pt]{article}
\usepackage[latin1]{inputenc}
\usepackage[english]{babel}
\usepackage[namelimits]{amsmath}
\usepackage{authblk}
\usepackage{amssymb}
\usepackage{amsmath}
\usepackage{amsthm}

\begin{document}
\title{{\bf Statistical description of massless excitations within a sphere with a linear equation of state and the dark energy case}}
\author[1,2,3]{S. Viaggiu\thanks{viaggiu@axp.mat.uniroma2.it and s.viaggiu@unimarconi.it}}
\affil[1]{Dipartimento di Matematica, Universit\`a di Roma ``Tor Vergata'', Via della Ricerca Scientifica, 1, I-00133 Roma, Italy.}
\affil[2]{Dipartimento di Fisica Nucleare, Subnucleare e delle Radiazioni, Universit\'a degli Studi Guglielmo Marconi, Via Plinio 44, I-00193 Rome, Italy.}
\affil[3]{INFN, Sezione di Napoli, Complesso Universitario di Monte S. Angelo,
	Via Cintia Edificio 6, 80126 Napoli, Italy.}

\date{\today}\maketitle
\begin{abstract}
In this paper we continue the investigations present in \cite{1}-\cite{3}. In particular, we extend the theorem proved in \cite{3}
to any massless excitation in a given spherical box. As a first interesting result, we show that it is possible, contrary to the black hole case
studied in detail in \cite{1,2,3}, to build macroscopic configurations with a dark energy equation of state. 
To this purpose, by requiring a stable configuration, a macroscopic dark fluid is obtained with an
internal energy $U$ scaling as the volume $V$, but with a fundamental correction looking like $\sim 1/R$ motivated by 
quantum fluctuations. Thanks to the proposition in section 3 
(and in \cite{3} for gravitons), one can depict the dark energy in terms of massless excitations with a discrete spectrum.
This fact open the possibility to test a possible physical mechanism converting usual radiation into dark energy in a macroscopic configuration, also in a cosmological context. In fact, for example, in a Friedmann flat universe with a cosmological constant
particles are marginally trapped at the Hubble horizon for any given comoving observer.

\end{abstract}
PACS Number(s): 05.20.-y, 04.30.-w, 04.70.-s, 04.60.-m

\section{Introduction}

At present time, no wideley accepted theory depicting gravity in a quantum regime is at our disposal. In particular, what is missing is
the presence of some phenomenological (Planckian) experimental fact leading the researchers to a physically and mathematically sound
formulation of a quantum gravity theory.\\ 
The study of free electromagnetic waves has played a
fundamental role for the quantization of the free electromagnetic field. In particular, 
the Planck formula for the radiation within a black body cavity, where photons are absorbed and emitted,
has marked the born of quantum mechanics.  
A particle (electron as an example) propagating in
a region of the order of its de Broglie wavelength can manifest quantum effects. 
In a similar way, 
since massless excitations provided by gravitons are widely believed to be the quanta of the gravitational field,
it is expected that the study of gravitational waves propagating in a 
finite size region can provide informations on possible manifestations of the quantum nature of the gravity. Moreover, the 
discovery \cite{4}  that black holes emit a strictly thermal radiation and are equipped with a non-vanishing entropy
\cite{5} proportional to their proper area  has represented a cornerstone for a possible formulation of a quantum gravity theory.
In particular, the features of gravitons, the search of the physical nature of the black hole entropy and the related logarithmic corrections represent still open issues (see for example \cite{6,7,8,9,10,11,12,13,14,15,16,17,18,19,20,21,22,23,24,25,26,27,28,29}).

Motivated by the reasonings above, in \cite{1} we have attempted to explain the degrees of freedom due to the non-vanishing black
hole entropy in terms of trapped gravitons inside the event horizon. The discrete spectrum is obtained by applying Dirichelet boundary conditions to a trapped gravitational radiation. As a result, a statistical interpretation for the black hole entropy can be 
obtained and after introducing some reasonings concerning a quantum spacetime \cite{30,31} and the related spacetime uncertainty relations
\cite{30,32,33}, corrections to the area law can arise \cite{2} by introducing at hand procedures. In both papers, gravitons satisfy a radiation equation of state. It is thus necessary to build a more physically sound procedure to obtain the black hole 
entropy corrections that are expected in a full quantum gravity (Planckian) regime. 
One can think that finite size effects due to the confinement procedure can change the equation of state of the trapped gravitons 
thus generating a $\gamma$-linear equation of state for the pressure $P$ and the energy density $\rho$ 
(i.e. $P=\gamma\rho c^2$, with $\gamma\in\mathbf{R}$). To this purpose, in \cite{3} a theorem has been proved where the 
spectrum of frequencies $\omega$ suitable for a $\gamma$-linear equation of state has been obtained. Quite remarkable, the modified equation of state for the trapped gravitons generates the well known logarithmic corrections, without assuming a particular quantum gravity proposal theory (loop, string...). Moreover, in
\cite{3} it has been shown that black holes with the interior filled with dark energy can be obtained only at microscopic Planckian size.

It can be argumented that, although gravitons can be trapped inside an event horizon of a black hole, the trapping procedure
in a more usual box (see \cite{1} and references therein) is not a simple task. To this purpose, we extend the treatment 
present in \cite{3} to  generic 
massless excitations that can be confined in a usual box. In particular, in order to give a physically sound statistical
description of dark energy, we are mainly interested to build macroscopic 
dark energy configurations.

In section 2 we present some statistical preliminaries. In section 3 we extend the results of \cite{3} to generic massless
excitations. In section 4 we study the dark energy case, while in section 5 the black hole case is analysed. 
In section 6 we compare our modeling with
other ones present in the literature in an astrophysical context.
Section 7 is devoted to conclusions and final considerations.

\section{Preliminaries: statistical mechanics of massless particles}

In \cite{1} we have shown that a suitable expression for the discrete spectrum for the allowed frequencies of the trapped gravitons in a spherical box of radial radius $R$ is given by:
\begin{equation}
{\omega}_{\ell n}\simeq\frac{c}{2R}\left(2+\ell+2n\right)\pi,\,\;\ell\geq 2,\;\;n\in\mathbb{N},
\label{1}
\end{equation}
where $\{\ell\}$ is the integer Legendre index with $\ell \geq 2$, 
$n$ is an integer quantum number and  $c$ denotes the speed of the light. Note the behavior $\sim\frac{1}{R}$ in (\ref{1}): in \cite{1}
we have shown that such a behavior for $\omega$ does imply a radiation equation of state ($\gamma=\frac{1}{3}$). In the following section we show that in a more general context, the behavior $\omega\sim \frac{1}{R}$ is a necessary and sufficient condition to have a radiation field. 

To start with, consider the case of a radiation field in a cubic box of proper size $L$. As well known, after taking the vector field
${\bf A}$ with ${\bf A}\sim e^{\imath({\bf k}.{\bf x}-\omega t)}$, where '.' denotes the scalar product, ${\bf x}$ Cartesian
coordiantes and $t$ the time coordinate, 
for the wave vector ${\bf k}$ we have ${\bf k}=\frac{2\pi}{L}(n_1,n_2,n_3)$ with
$\{n_i\}\in\mathbf{Z}$. Hence, if we consider a radiation made of massless excitations on a spherical box of radial radius 
$R$, for the angular frequency $\omega$ and the energy $E_0$ we can write:
\begin{equation}
{\omega}_0=a\frac{c{\bf n}}{R},\;\;\;E_0=\hbar{\omega}_0,\;\;{\bf n}\in\mathbb{N},
\label{2}
\end{equation} 
with $a\in\mathbf{R^+}$ left unspecified. As usual \cite{1}, we can consider an ensemble of $N$ massless
excitations. The partition function 
$Z_T$ is given by $Z_T={(Z^{(0)})}^N$ with
\begin{equation}
Z^{(0)}=\sum_{n=0}^{\infty}e^{-\beta\hbar{\omega}_0}=\frac{1}{1-e^{-\frac{ac\hbar\beta}{R}}},
\label{3}
\end{equation}
where $\beta=\frac{1}{K_BT}$, $K_B$ denotes the Boltzmann constant and $T$ the thermodynamical temperature.
For the free energy $F_T$ we have $F_T=-K_BT\ln(Z_T)$ and thus for the internal energy $U^{(0)}=-{(\ln(Z_T))}_{,\beta}$
we obtain:
\begin{equation}
U^{(0)}=\frac{ac\hbar N}{R\left(e^{\frac{ac\hbar\beta}{R}}-1\right)}.
\label{4}
\end{equation}
After using the thermodynamical volume $V$ \cite{34} as $V=\frac{4\pi R^3}{3}$, for the pressure 
$P^{(0)}$ we obtain $P^{(0)}V=\frac{U^{(0)}}{3}$, suitable for a radiation field.

As customary \cite{35}, in a general relativistic classical context, we may think to a spherical box with radial radius $R$ with interior metric in the usual 
coordinates $(t,r,\theta,\phi)$ given by
\begin{equation}
ds^2=-e^{2f(r)} dt^2+\frac{dr^2}{\left(1-\frac{2 G E(r)}{c^4 r}\right)}+r^2\left(d\theta^2+\sin^2(\theta)d\phi^2\right),
\label{5}
\end{equation}
where $E/c^2$ denotes the mass-energy of the radiation inside:
\begin{eqnarray}
& & E(r)=4\pi c^2\int_0^r \rho(r^{\prime}){r^{\prime}}^2dr^{\prime},\label{6}\\
& & \frac{df}{dr}=\frac{\frac{GE(r)}{c^4}+\frac{4\pi G}{c^4} r^3 P(r)}{r[r-2\frac{G E(r)}{c^4}]}.\label{7}
\end{eqnarray}
The matching condition ensuring the continuity of the first fundamental form at $r= R$ requires $M=E(R)/c^2$.
The continuity of the second fundamental form, if required, avoids a boundary surface layer at $r=R$.  
Note that in this way $M$ is the ADM mass (total energy) of the spacetime.
As well known, the proper mass $M_p$ is given by
$M_p=4\pi c^2\int_0^R\rho(r)dV^{(3)}$, where $V^{(3)}$ is the proper volume inside the spherical mass-energy distribution, with
$U_B=(M_p-M)c^2>0$ denoting the binding energy. As a result of these considerations, in a thermodynamical context we identify the ADM
(not the proper one $M_p$) mass-energy $E(R)=Mc^2$
with the internal energy of the spherical configuration, i.e. $U=Mc^2$. This is in agreement with the interpretation of the internal energy as the energy necessary to create a given system (see for example \cite{1}). Hence, in the following sections, to obtain a
suitable expression for the internal energy $U$, we integrate the mass-energy density $\rho$ over the 'thermodynamical'
(see \cite{34} in a cosmological context) volume element $4\pi r^2 dr$ rather than over the proper one $dV^{(3)}$.

\section{Massless excitations with a linear equation of state}

In the section above we have shown that angular frequencies looking like $\sim\frac{1}{R}$ lead to a radiation field with 
$PV=\frac{U}{3}$. In this section we show that, using a very similar version of the proposition in \cite{3}, this does
happen in a more general context. To start with, consider $N$ generic massless excitations 
enclosed in a spherical box of proper areal radius
$R$ with angular frequency ${\omega}_0$:
\begin{equation}
{\omega}_0=\frac{\phi(R)}{N}.
\label{8} 
\end{equation}
For the internal energy $U_0$ we obtain $U_0=\hbar\phi(R)$.\\ 
After using the usual relations $F_{0T}=-K_B T\ln(Z_{0T})$ and
$\frac{\partial F_{0T}}{\partial V}=-P^{(0)}$ with $Z_{(0)T}=e^{-\hbar\beta\phi}$ and $V=4\pi R^3/3$, we obtain
\begin{equation}
P^{(0)}V=-\frac{\hbar R}{3}\frac{d\phi(R)}{dR}-\frac{\hbar}{3}\phi(R)+\frac{U^{(0)}}{3}.
\label{9}
\end{equation}
After imposing the constraint $P^{(0)}V=\frac{U^{(0)}}{3}$ we have:
\begin{equation}
R\frac{d\phi(R)}{dR}+\Phi(R)=0,
\label{10}
\end{equation}
with the general solution $\phi(R)=\frac{k}{R},\;k\in\mathbf{R^+}$. Hence, a necessary and sufficient condition to have a free radiation field in a spherical box is that the quanta must oscillate at frequencies proportional to $1/R$.

As shown in \cite{1}, a massless field made of gravitons (\ref{1}) can explain the BH entropy formula. However, as well known, 
logarithmic corrections to the BH formula are expected. To obtain these corrections, taking the radiation equation of state (\ref{1})
held fixed, one can adopt euristic \cite{2} arguments dictated by statistical fluctuations. 

A more rigorous way to obtain logarithmic corrections is to modify the formula (\ref{1}), i.e. to suppose that quantum effects due to the 
finiteness of the areal radius $R$ can modify the radiation equation of state in a $\gamma$ linear one $PV=\gamma U$.
As a consequence, logarithmic corrections to the BH entropy do arise. Moreover, in \cite{3} we have shown that, in particular, dark energy-like configurarions for a BH are only possible at Planckian scales, and as a result macroscopic
black holes with an interior made of dark energy-like gravitons are forbidden. It is thus interesting to ask if macroscopic configurations (different 
from the black hole ones) can be obtained with a dark-energy like equation of state. This is a fundamental task in order to explain the 
dark energy in the form of the cosmological constant $\Lambda$ from a quantum-statistical point of view, i.e. in order to explain the physical nature of the dark energy.\\
To start with, we easily extend the theorem \cite{3} for any massless excitation:

{\bf Proposition:} {\it Let ${\omega}_0$, given by the (\ref{2}), denote the angular frequency of a radiation field composed of $N$ 
massless excitations enclosed in a spherical box of proper areal radius $R$, then the ones 
with angular frequency $\omega={\omega}_0+\frac{\Phi(R)}{N}$ and given internal energy $U(R)$ have a linear equation of state
$PV=\gamma U$ if and only if
the function $\frac{\phi(R)}{N}$ satisfies the following equation} 
\begin{equation}
\hbar\left[R\;\frac{d\Phi}{dR}+\Phi(R)\right]=U(R)(1-3\gamma), \label{11}
\end{equation}
{\it together with the condition}
\begin{equation}
U-\hbar\;\phi(R) > 0.
\label{12}
\end{equation}
\begin{proof}
We have
\begin{equation}
U=U^{(0)}+\hbar\;\Phi(R)
\label{13}
\end{equation}	
with $U^{(0)}$ given by (\ref{4}): condition (\ref{12}) is thus a consequence of the fact that from (\ref{4}) we have $U^{(0)}>0$.\\
Instead of the (\ref{9}) we have
\begin{equation}
PV=-\frac{\hbar R}{3}\frac{d\Phi(R)}{dR}-\frac{\hbar}{3}\Phi(R)+\frac{U}{3},
\label{14}
\end{equation}
and the (\ref{11}) holds after imposing $PV=\gamma U$. 
\end{proof}
Equation (\ref{11}) implies that for a radiation field with $\gamma=\frac{1}{3}$, we have $\omega\sim\frac{1}{R}$, as depicted by equation (\ref{10}) that obviously is nothing else but the homogeneous part of the (\ref{11}). From the calculations above it follows that
the proved theorem must hold starting from any radition field (gravitons, photons..) trapped in a certain finite spherical box.
The starting ingredient is the behavior ${\omega}^{(0)}\sim\frac{1}{R}$.\\
Also note that the added frequency term $\Phi(R)/N$ is independent on some quantum number, for example it is not 
multiplied by some integer $I\in\mathbb{N}$. This choice is physically motivated by the fact that in our background the term
$\Phi(R)$ is generated by quantum fluctuations from finite size effects that on general grounds are expected to generate an 
added frequency term dependending only on the radius $R$ and on the collective presence of $N$ massless excitations.

The next step is to study the possible expressions for the total internal energy $U$\footnote{It is important to note that we fix
the expression for the total internal energy $U$ and not the one for $U^{(0)}$: in practice we calculate the total frequency 
$\omega$ needed to obtain a $\gamma$-linear equation of state for a given expression of $U(R)$.} 
as a function of $R$. As shown in \cite{2,3}, we could consider
energy-density looking like\footnote{In practice for massless excitations
we consider $\rho c^2=\frac{dE}{dV}$, see \cite{2,3}.}
$\rho=Br^w,\;B\in\mathbf{R^+}$. For the statibility of the configuration, we require $w\leq 0$ and
$w>-3$ for integrability.
We can take $U=\frac{c^4}{2G}K_w{R^{w+3}},\;K_w\in\mathbf{R^+}$. The black hole case can be obtained with $w=-2$.
We can thus solve equation (\ref{11}). Since the homogeneous solution is proportional by a constant to $1/R$ and requiring that for $\Phi(R)=0$ (radiation) the solution is provided by
(\ref{2}), we obtain $\hbar\Phi(R)=\frac{(1-3\gamma)c^4 K_w}{2G (w+4)}R^{w+3}$. Condition (\ref{12}) it gives
\begin{equation}
\gamma > \frac{-3-w}{3},\;\;\;w\in(-3,0].
\label{15}
\end{equation}
For a black hole ($w=-2$) condition (\ref{15}) means that the active gravitational mass must be positive.\\
In order to obtain a statistical description of the cosmological constant, we must consider the case with 
$\rho=B$, i.e. $w=0$: in this case we have $\gamma >-1$. As a consequence, the case suitable for the cosmological constant
$\gamma=-1$ is forbidden. However, it is also remarkable that macroscopic configurations vith $\gamma\in(-1,\frac{1}{3}]$
are allowed\footnote{In a cosmological context, Friedmann cosmologies with $\gamma<-\frac{1}{3}$ lead to accelerating universes.}.\\
It is interesting to note that, according to the reasonings present in \cite{36}, the dark energy configuration can be obtained in the 
extreeme limit for $T\rightarrow 0$, i.e. $U^{(0)}\rightarrow 0$.
By inspection of (\ref{4}), we see that the feasibility of this limit depends on the behavior of
$N$. In practice, bosons must condensate in the state at $T=0$. Apart from this interesting possibility
that can be matter of future investigations, we are interested, in line
with the research in \cite{2,3}, to find suitable physically motivated expressions for the internal energy $U(R)$ that are capable to 
depict macroscopic finite configurations supporting a dark energy-like content within. This task will be accomplished in the next section.

\section{Building macroscopic dark energy spheres}

In order to modelise dark energy expressed in term of a positive cosmological constant $\Lambda$, we consider the case $\rho=B$.
In terms of the internal energy we have
\begin{equation}
U(R)=\frac{c^4}{2G}K_0 R^3,\;\;\;B=\frac{3c^2 K_0}{8\pi G}.
\label{16}
\end{equation}
This case has been partially analysed in \cite{3}. In order to obtain a dark energy equation of state with\footnote{Here we used the
expression $\frac{c^2\Lambda}{8\pi G}={\rho}_{\Lambda}$ to mimic the cosmological case, but our reasonings in this paper concern a static 
background.} we can write $\Lambda=3K_0$. As shown in \cite{3}, to get the equation of state with $\gamma=-1$ 
(or with $\gamma<-\frac{1}{3}$), Planckian effects must been introduced. In particular, if we suppose possible quantum corrections due to 
finite size effects, one can assume a quantum spacetime structure \cite{30,31}: 
physically motivated spacetime uncertainty relations can be obtained \cite{30,32} (see also \cite{33} for the cosmological case) that
after assuming spherical symmetry (the same of the spherical box) become
\begin{equation}
{\Delta}_{\omega} R\geq s L_P,\;\;\;c{\Delta}_{\omega} t{\Delta}_{\omega} R\geq s^2L_P^2,\;\;s\sim 1,
\label{21}
\end{equation}
where $\omega$ denotes a generic quantum state and $L_P$ the Planck length. From the time-energy uncertainty relation we have
${\Delta}_{\omega}E{\Delta}_{\omega} t\geq\frac{\hbar}{2}$. The minimal uncertainty 
(i.e. maximal localizing gaussian states \cite{30} satisfying the (\ref{21})) is obtained
with states such that $c{\Delta}_{\omega} t\sim {\Delta}_{\omega} R$ and as a result
${\Delta}_{\omega} E\sim 1/R$. As a consequence of these reasonings, a physically motivated expression for the
total modified internal energy
$U$ is
\begin{equation}
U(R)=\frac{K_0 c^4}{2G}R^3+\frac{K_1 c^4 }{2GR},\;\;\;\;\{K_1\}\in{\mathbf{R}}.
\label{22}
\end{equation}
Note that in the (\ref{22}) we have left open the possibility that $K_1<0$. After inserting the (\ref{22}) in 
(\ref{11}) with $\gamma=-1$, for $\Phi(R)$ we obtain
\begin{equation}
\hbar\Phi(R)=\frac{K_0 c^4}{2G}R^3+\frac{K_1 c^4}{2GR}\ln\left(\frac{R}{R_0}\right),
\label{23}
\end{equation}
where $R_0$ is an integration constant with the order of magnitude to be discussed. 
Condition (\ref{12}) thus becomes
\begin{equation}
\frac{K_1 c^4}{2GR}-\frac{K_1 c^4}{2GR}\ln\left(\frac{R}{R_0}\right)>0.
\label{24}
\end{equation}
It is worth to be noticed that only in the dark energy case with $\gamma=-1$, terms proportional to $K_0$ in 
(\ref{22}) and (\ref{23}) have the same coefficient and as a result condition (\ref{24}) is independent from $K_0$: this 
fact further indicates that the case $\gamma=-1$ is indeed a very exceptional intriguing case.\\
Equation (\ref{22}) means that the classical expression for the internal energy  $U\sim R^3$ 
(i.e. $\rho\in\mathbf{R^+}$) is corrected (dressed) by quantum effects.\\
From time-energy uncertainty relation combinated with (\ref{21}), we expect that  $|K_1|\sim\L_P^2$.  
Note that the term proportional to $K_0$ in (\ref{23}) is due to the term $\sim R^3$ in (\ref{22}), while
the one proportional to $K_1$ is due to the quantum correction in (\ref{22}): we expect that at the order of the Planck length
the two contributions are at least comparable, i.e. $|K_1|\geq s^4 K_0 L_P^4\rightarrow K_0\leq\sim\frac{1}{L_P^2}$,
with $s$ of the order of unity. Conversely,
for $R>>L_P$ the term proportional to $K_0$ will be the dominant one.\\
Moreover, for the 
physically motivated reasoning above, we expect that the term proportional to $K_1$, i.e.
$\frac{c^4}{2GR}\ln\left(\frac{R}{R_0}\right)$ in (\ref{23})
has a maximum near the Planck regime: since this term has an absolute maximum for $R=e R_0$, we thus expect that 
$R_0=sL_P$.\\
In the next subsections we analyse both cases with $K_1>0$ and $K_1<0$.

\subsection{The case with $K_1>0$}
For this case, from the (\ref{22}) obviously we have $U(R)>0$. Condition (\ref{24}) with $R_0=eL_P$ is satisfied for 
$R<esL_P$. Hence, according to the partial study of \cite{3}, a maximum allowed value for $R$ of the order of the 
Planck length does arise and as a consequence only microscopic configurations can be obtained. 
We have restricted our considerations to the 
important case of the cosmological constant, but similar restrictions can be found for $\gamma<-\frac{1}{3}$.

\subsection{The case with $K_1<0$}

In the case $K_1<0$, obviously condition (\ref{24}) it gives $R>es L_P$. However, further 
conditions must be imposed. First of all, the internal energy is required to be positive: $U(R)>0$:
this condition does imply that $K_0 R^4-|K_1|>0$, i.e.
\begin{equation}
R>{\left(\frac{|K_1|}{K_0}\right)}^{\frac{1}{4}}.
\label{25}
\end{equation}
Since from the computations below equation (\ref{24}) we have $|K_1|\geq s^4 K_0 L_P^4$,
the (\ref{25}) implies that $R>s L_P$ that in turn is implicated by $R>es L_P$.
As a further resctrictive condition that can be imposed \footnote{This is a sufficient but not necessary condition 
ensuring that $U^{(0)}+\hbar\Phi(R)>0$, with $U^{(0)}$ given by (\ref{4}).} is $\Phi(R)>0$:
\begin{equation}
K_0 R^4-|K_1|\ln\left(\frac{R}{sL_P}\right) > 0.
\label{26}
\end{equation}
After denoting $\frac{K_0}{|K_1|}=\frac{\delta}{L_P^4}$ with $\delta<1$ and $\frac{R}{sL_P}=W$, the (\ref{26})
becomes $\delta W^4-\ln(W)=H(W)>0$ with $W\geq 1$. The function $H(W)$ has an absolute minimum at
$W_m={\left(\frac{1}{4\delta}\right)}^{\frac{1}{4}}$: condition (\ref{26}) is thus satisfied provided that
$H(W_m)>0\rightarrow 1+\ln(4\delta)>0$ and as a consequence we obtain the solution 
$\delta\in\left(\frac{1}{4e},\;1\right)$.\\
These computations show that also after imposing restrictive conditions, we can build dark energy 
macroscopic configurations. 

Summarizing, classically we can obtain macroscopic dark energy-like configurations with $\rho=B\in\mathbf{R^+}$ only in the range
$\gamma\in\left(-1,\;-\frac{1}{3}\right)$. The cosmological constant case $\Lambda$, with energy density ${\rho}_{\Lambda}$ given by
${\rho}_{\Lambda}=\frac{c^2\Lambda}{8\pi G}$, together with $k_0=\frac{\Lambda}{3}$, can be obtained only when Planckian
effects are considered. These quantum effects can be phenomenologically understood as producing an extra term to the energy density
$U/c^2$ looking like $K_1/R$. For $K_1>0$ only Planckian configurations are possible, while with 
$K_1<0$ also macroscopic configurations with $R>>L_P$ are available.\\
It is thus interesting to ask if this extra term can be depicted in terms of a modified (constant) energy density. The answer is  affirmative. In fact, we can consider the density ${\rho}_{\Lambda}=\frac{c^2\Lambda}{8\pi G}=B$ as a 'bare' mass-energy
density. After that quantum effects are taken into account, we have a 'dressed' cosmological constant
$\overline{\Lambda}$ with a 'dressed' constant energy-density ${\rho}_{\overline{\Lambda}}=\overline{B}$ with
\begin{equation}
\overline{B}=B+\frac{3c^2 K_1}{8\pi G R^4}=\frac{c^2\overline{\Lambda}}{8\pi G}.
\label{27}
\end{equation}
Note that equation (\ref{27}) indicates that the dressed cosmological constant $\overline{\Lambda}$ is a function of the
proper areal radius $R$.\\
The considerations above clearly show that, in order to obtain a physically motivated statistical description of the cosmological
constant $\Lambda$, quantum reasonings motivated by a possible quantum nature of the 
spacetime, at least at a 'phenomenological' level, may be of great importance: this fact is 
a strong indication of the quantum origin of $\Lambda$.

\section{The black hole case}

In the black hole case, as analysed in \cite{3} with $\gamma=-1$, the internal energy is (we have rescaled $K_1$ of section 4
by the Planck length $L_P$)
\begin{equation}
U(R)=\frac{c^4 R}{2G}+\frac{K c^4 L_P^2}{2GR}.
\label{28}
\end{equation}
The case with $K>0$ has been studied in \cite{3}. As a result of this study, only dark energy configurations of Planckian size are
possible. However, we can ask if this result is changed by considering $K<0$ as done in section 4. First of all, the condition
$U(R)>0$ implies that $R>\sqrt{|K|} L_P$. After denoting again with $\frac{R}{sL_P}=W$, the condition 
(\ref{12}) becomes:
\begin{equation}
-W^2-|K|+4|K|\ln\left(\frac{W}{s}\right)= F(W)> 0.
\label{29}
\end{equation}
Function $F(W)$ has a maximum at $W_m=\sqrt{2|K|}$ and is positive $\forall W\in(W_a,W_b)$ with $W_m\in(W_a,W_b)$.
After imposing $F(W_m)>0$, we obtain $|K| > \frac{s^2}{2}e^{\frac{3}{4}}$. 
Thanks to the discussion of section 4 regarding $K_1$ we expect
$s$ and $K$ of the order of unity. As an example, for $s\sim 1$, we obtain: for $|K|=3\rightarrow$ $W_a\simeq 1.6, W_b\simeq 3.4$,
while for $|K|=10\rightarrow$ $W_a\simeq 1.3, W_b\simeq 8.8$. These computations clearly show that, according to \cite{3}, black hole with 
interior composed of a cosmological-like equation of state can only exist at Planckian scales.

\section{Our method in an astrophysical context}

In the section above we have considered in detail the case of dark energy expressed in terms of the cosmological constant
$\Lambda$. However, note that several
self-gravitating physical systems could be addressed with the same technique depicted by formula (\ref{11}).

The presence of the term $\Phi(R)$ is motivated by quantum fluctuations. It is thus interesting to  compare our
approach with other ones where modifications to the Einstein's equations are proposed. As a first example, in \cite{r1}, the
authors considered the modification to the Newtonian potential in the $f(R)$ gravity context in order to study a statistical system of 
$N$ particles representing galaxies. As a result they obtain a modification of the usual internal energy in the form 
$U=\frac{3Nk_BT}{2}(1-2\beta)$, where $\beta$ is the clustering parameter with embedded the $f(R)$ gravity contribution.
Moreover, the equation of state becomes $PV=NK_B T(1-\beta)$.
In this paper we considered massless excitations and the expression for $U$ for large radius configurations is
$U=NK_BT+\hbar\Phi$, while in \cite{r1} massive particle are considered. In any case, the physical interesting analogy is that also in our case the quantum fluctuations induced from finite size effects act modifying
the internal energy by an added term provided by $\Phi$. It is thus interesting to ask what happens if the calculations present in
\cite{r1} are redone for a massless cluster of particles in the $f(R)$ context. This could be matter for future investigations.\\
Also note that, in order to consider more general equations of state than
the $\gamma$ linear one, we can easily modify the formula (\ref{11}) in the following way: 
\begin{equation}
\hbar\left[R\;\frac{d\Phi}{dR}+\Phi(R)\right]=U(R)-3\left[F(U)+G(R)\right], \label{r11}
\end{equation}
with the general equation of state $PV=F(U)+G(R)$: after posing $\rho=U/V$, we can obtain a general barotropic equation of state with 
a further term added depending from $R$, i.e.
$P=Q(\rho)+G(R)/V$. 
A possible application of the (\ref{r11}) can be in the modeling of the core of neutron 
or very massive stars, where a more complicated expression for $F(U)$ is expected than the one given by the $\gamma$ linear case.
In a cosmological context for the dark energy, insted of the usual equation of state with $\gamma=-1$ more general ones can be considered
(see for example \cite{r2} and \cite{r3}). As an example, in \cite{r2} the equation of state 
$p=-\rho-A{\rho}^{\alpha}-BH^{2\beta}$, with $\{A,\alpha,\beta\}\in\mathbf{R}$ and $H$ the Hubble parameter, is proposed. Also the 
so called quadratic one \cite{rr2} is studied in \cite{r2}. All these proposed equations of state can be easily obtained within our approach and the theorem proved with the (\ref{11}) can be generalised with the formula (\ref{r11}) instead of the (\ref{11}).
To obtain the case in \cite{r3}, we can choose $F(U)=0$ and $G(R)\sim\ln(V)=\ln(R^3)$.

The reasonings above show that our approach is rather general and can be an useful tool to investigate several interesting 
self-gravitating systems.

\section{Conlusions and final remarks}

In paper \cite{3} we ask to the following question: what is the discrete spectrum that trapped massless gravitons 
within a confining box should have in order to have a $\gamma$-linear equation of state ? In other words, we start from a given 
expression for the internal energy $U(R)$ suitable for a given chosen 
physical model and we found the frequencies spectrum that massless 
gravitons must have in order to give a $\gamma$-linear equation of state. In turn, thanks to finite size effects,
we depict at a phenomenological level a possible quantum mechanism transforming a radiation field in a dark energy one. 
Althought 
the trapping procedure for gravitons in a box is 
a debated and controversial question (see the introduction of \cite{1} and references therein), we expect that gravitons can be trapped within the event horizon of a black hole. In this paper we extend the results present in \cite{3} to a generic starting ensemble of $N$
massless excitations with angular frequency ${\omega}_0$ given by (\ref{2})
within a given spherical box, not necessarily a black hole. We have found that, starting with a radiation field of massless excitations, there exists a smooth differentiable
function $\Phi(R)$ such that the freqeuncy $\omega={\omega}_0+\frac{\Phi(R)}{N}$ depicts a fluid with a
$\gamma$-linear equation of state and with a given internal energy $U(R)$, 
provided that equation (\ref{11}) with condition (\ref{12}) are satisfied. From a physical point of view, we may suppose that quantum gravity effects due to finite size effects change the frequency spectrum of the radiation inside by a 
quantity depicted by $\Phi(R)$. In this regard, $\hbar\Phi(R)$ is the energy added to make this change in frequency and the term $\Phi(R)/N$ is the 'mean' value received from any massless excitation. In practice, massless excitations acquire an 'effective' mass term
depending on $\Phi(R)/N$: for large values of the areal radius $R$, this effective mass is expected to be negligible.\\
Our approach is 'phenomenological' and does not require an underlying quantum gravity theory.

In this paper we focused our attention to the study of dark energy configurations with constant positive energy 
density within a box, i.e. the case of
a positive cosmological constant. In section 4 we have shown that also under very restrictive conditions, it is possible to obtain macroscopic configurations with massless excitations where a statistical description of the cosmological
constant emerges, provided that corrections to the internal energy loooking like $1/R$ and motivated by quantum 
fluctuations are
introduced. It is worth to be noticed that our treatment does not require an underlying quantum gravity theory. In this regard, our approach is 'phenomenological' and the physical effects depicted are expected, on general grounds, to be included in a more general
quantum gravity theory.\\
From the point of view of general relativity, we may think to a sphere filled with a positive cosmological constant (de Sitter) given by
\begin{equation}
ds^2=-\left(1-\frac{{\overline{\Lambda}} r^2}{3}\right) dt^2+
\frac{dr^2}{\left(1-\frac{{\overline{\Lambda}} r^2}{3}\right)}+r^2\left(d\theta^2+\sin^2(\theta)d\phi^2\right).
\label{30}
\end{equation}
The exterior metric for $r>R$ is given by the Schwarzschild one. Continuity of the first fundamental form requires that
the ADM mass $M$ must satisfy the equality $M=\frac{c^2R^3\overline{\Lambda}}{6G}$, in agreement with the computations of section 4.
Obviously, the continuity of the second fundamental form cannot be accomplished and as a result we have a distributional source on 
$r=R$. This distributional source
can be composed, for example, of anti de Sitter ($\Lambda<0$) cosmological constant acting as a confining material\footnote{To this purpose, the reader can read the introduction of \cite{1} and references therein.}. In practice, also a 
cavity trapping the massless excitations inside can work. Details of this cavity are not relevant for the results of this paper.

As a final consideration, note that in this paper we considered an expression for $U(R)$ looking like $U(R)\sim R^3$: this behavior lead to an 
energy density $\rho=B$. Unfortunately, this behavior, after solving the (\ref{11}), it gives models with $\gamma>-1$,
thus excluding the cosmological constant case with constant energy-density $Bc^2$ and $\gamma=-1$.
To obtain the important case of a cosmological constant, motivated by possible effects expected at Planckian scales,
we must add a term proportional to $1/R$ to $U(R)$ (expression (\ref{22})). As a result, macroscopic configurations are also possible
under rescrictive hypothesis. In formulas, the internal energy $U(R)$ (\ref{22}) does imply an effective (dressed) 
cosmological constant $\overline{\Lambda}$, namely equation  (\ref{27}), given by
\begin{equation}
\overline{\Lambda}(R) 
\label{31}={\Lambda}_0+\frac{3K_1}{R^4},\;\;\;{\Lambda}_0=\frac{8\pi G B}{c^2}.
\end{equation}
Physically, $\overline{\Lambda}$ can be interpreted as an 'interacting' cosmological constant: after that Planckian effects are taken into account, the bare cosmological constant ${\Lambda}_0$ is dressed by quantum fluctuations. This interpretation is reminescent of
renormalization group techniques applied to the cosmological constant. In fact, as shown in 
\cite{37}, after applying renormalization group techniques to a Friedmann flat universe equipped with a non-vanishing positive
cosmological constant ${\Lambda}_0$, an effective running cosmological constant does appear $\Lambda(H)$, where 
$H(t)$ is the Hubble rate: 
\begin{equation}
\Lambda(H)=c_0+3{\nu}_1H^2+3{\nu}_2 H^4,
\label{32}
\end{equation}
where $\{c_0, {\nu}_1,{\nu}_2\}$ are constants characterazing quantum-particle calculations. In terms of the Hubble radius
(apparent horizon) $R_h=\frac{c}{H}$, the (\ref{32}) becomes
\begin{equation}
\Lambda(R_h)=c_0+\frac{3c^2{\nu}_1}{R_h^2}+\frac{3c^4{\nu}_2}{R_h^4}.
\label{33}
\end{equation}
Although formulas (\ref{31}) and (\ref{33}) have been obtained in different contexts (static and Friedmann respectively), the analogy 
emerges. In (\ref{31}) the term $1/R^4$ is a consequence of the term $\sim 1/R$ in $U$, while the term
${\Lambda}_0$ is due to the one $\sim R^3$. Suppose now that the machinery
used in this paper is extended sic simpliciter to a Friedmann flat universe. If we assume as internal energy the Misner-Sharp \cite{38}
mass-energy $M_{ms}c^2$, then exactly at the Hubble horizon \cite{39,40,41} $R_h$ we have 
$M_{ms}c^2=U(R_h)=\frac{c^4 R_h}{2G}$ and as a result the term $\sim 1/R_h^2$ arises since ${\rho}_{\Lambda}\sim 1/R_h^2$ with
${\Lambda}\sim 1/R_h^2$: in this way we obtain a cosmological constant of the same order of magnitude of the one actually observed.
This is certainly an encouraging starting point.


\begin{thebibliography}{0}
\bibitem{1}S. Viaggiu,  {\it Physica A} {\bf 473} (412 (2017). 
\bibitem{2}S. Viaggiu, {\it Physica A} {\bf 488} 72 (2017). 
\bibitem{3}S. Viaggiu,  {\it Int. Jour. Mod. Phys. D} {\bf 27} 1850061 (2018).
\bibitem{4}S. Hawking, {\it Commun. Math. Phys.} {\bf 43} (1975) 199.
\bibitem{5}L.D. Bekenstein, {\it Phys. Rev. D} {\bf 23} 287 (1981).	
\bibitem{6}L. Smolin, {\it Gen. Rel. Gravit.} {\bf 16}  208 (1985).
\bibitem{7}L. Smolin, {\it Gen. Rel. Gravit.} {\bf 17} 417 (1985).
\bibitem{8}D. Garfinkle and R.M. Wald, {\it  Gen. Rel. Gravit.} {\bf 17} 461   (1985).
\bibitem{9}T. Padmanabhan and T.P. Singh, {\it Class. Quantum Grav.} {\bf 20} 4419 (2003).
\bibitem{10}E. Bianchi  arXiv:1211.0522, preprint 
\bibitem{11}C. Corda, {\it Eur. Phys. J. C} {\bf 73} 2665 (2013).
\bibitem{12}C. Corda, {\it Class. Quantum Grav.} {\bf 32} 195007 (2013).
\bibitem{13}L. Susskind and G. Uglum, {\it Phys. Rev. D} {\bf 50} 2700 (2013).
\bibitem{14}L. Bombelli, R. Koul, J. Lee and R. Sorkin, {\it Phys. Rev. D} {\bf 34} 373 (1986).
\bibitem{15}V. Frolov and I.  Novikok, {\it Phys. Rev. D} {\bf 48} 4545 (1993).
\bibitem{16}M. Srednicki, {\it Phys. Rev. Lett.} {\bf 71} 666 (1993).
\bibitem{17}R.M. Wald, {\it Phys. Rev. D} {\bf 48} 3427 (1993).
\bibitem{18}A. Strominger  and C. Vafa, {\it Phys. Lett B} {\bf 104} 379 (1996).
\bibitem{19}V.P. Frolov, D.V. Fursaev and A.I. Zelnikov, {\it Nucl. Phys. B} {\bf 486} 339 (1997).
\bibitem{20}C. Rovelli, {\it Phys. Rev. Lett.} {\bf 77} 3288 (1996).
\bibitem{21}S. Carlip, {\it Class Quantum Grav.}{\bf 16} 3327 (1999).
\bibitem{22}L. Susskind, {\it Jour. Math. Phys.} {\bf 36} 6377 (1995).
\bibitem{23}T. Jacobson, {\it Phys. Rev. Lett.} {\bf 75} 1260 (1995). 
\bibitem{24}S. Carlip, {\it Class Quantum Grav.} {\bf 17} 4175 (2000).
\bibitem{25}R. Dijkgraaf, H. Verlinde and E. Verlinde, {\it Nucl. Phys. B} {\bf 371} 269 (1992).
\bibitem{26}M. Cavagli\'a and C. Ungarelli, {\it Phys. Rev. D} {\bf 61} 064019 (2000). 
\bibitem{27}M. Cavagli\'a and A. Fabbri, {\it Phys. Rev. D} {\bf 65} 044012 (2002).
\bibitem{28}D.N. Page, {\it New J. Phys.} {\bf 7} 20 (2005).
\bibitem{29}E.A. Novikov, {\it Mod. Phys. Lett. A} {\bf 31} 1650092 (2016). 
\bibitem{30}S. Doplicher, K. Fredenhagen and J.E. Roberts,  {\it Comm. Math. Phys.} {\bf 172} 187 (1995). 
\bibitem{31}D. Bahns, S. Doplicher, G. Morsella  and G. Piacitelli, 
{\it Advances in Algebraic Quantum Field Theory, 289-330 Springer} (2015).
\bibitem{32}L. Tomassini and S. Viaggiu,  {\it Class. Quantum Grav.} {\bf 28} 075001 (2011). 
\bibitem{33}L. Tomassini and S. Viaggiu, {\it Class. Quantum Grav.} {\bf 31} 185001  (2014). 
\bibitem{34}D. Kastor, S. Ray and J. Traschen, {\it Class. Quantum Grav.} {\bf 26} 195011 (2009).
\bibitem{35}R.W. Wald, {\it General Relativity}, The university of Chicago Press, Ltd.,London (1984)
\bibitem{36}S. Viaggiu, {\it Int. J. Mod. Phys. D} {\bf 3} 1650033 (2016). 
\bibitem{r1}S. Capozziello, M. Faizal, M. Hameeda, B. Pourhassan, V. Salzano and S. Upadhyay, {\it Mont. Not. Roy. Astron. Soc.}
{\bf 474} 2430 (2018).
\bibitem{r2}S. Capozziello, V.F. Cardone, E. Elizalde, S. Nojiri and S.D. Odintsov, {\it Phys. Rev. D} {\bf 73} 043512 (2006).
\bibitem{r3}S. Capozziello, R. D'Agostino and O. Luongo, {\it Phys. Dark Universe} {\bf 20} 1 (2018).
\bibitem{rr2}B. Ratra and P.J.E. Peebles, {\it Phys. Rev. D} {\bf 37} 3406 (1988).
\bibitem{37}S. Basilakos, J.A.S. Lima and J. Sol\'a, {\it Int. J. Mod. Phys. D} {\bf 22} 1342008 (2013).
\bibitem{38}C.W. Misner and D.H. Sharp, {\it Phys. Rev.} {\bf 136} B571 (1964).
\bibitem{39}S. Viaggiu,  {\it Gen. Relativ. Gravit.} {\bf 48}:100 (2016). 
\bibitem{40}S. Viaggiu, {\it Gen. Relativ. Gravit.} {\bf 47}: (2015).
\bibitem{41}S. Viaggiu, {\it Mod. Phys. Lett. A} {\bf 4} 1650016 (2016). 
\end{thebibliography}
\end{document}